\documentclass[a4paper,11pt]{article}
\usepackage{pos}

\title{A novel approach to semileptonic heavy-to-light $B$ decays through the Dispersive Matrix method}
\ShortTitle{A novel approach to semileptonic heavy-to-light $B$ decays}

\author[a]{G. Martinelli}
\author[b]{S. Simula}
\author*[c]{L. Vittorio}

\affiliation[a]{Physics Department and INFN Sezione di Roma La Sapienza,\\
  Piazzale Aldo Moro 5, 00185 Roma, Italy}
  
\affiliation[b]{Istituto Nazionale di Fisica Nucleare, Sezione di Roma Tre,\\
Via della Vasca Navale 84, I-00146 Rome, Italy}

\affiliation[c]{LAPTh, Universit\'e Savoie Mont-Blanc and CNRS, Annecy, France}

\emailAdd{guido.martinelli@roma1.infn.it}
\emailAdd{silvano.simula@roma3.infn.it}
\emailAdd{ludovico.vittorio@lapth.cnrs.fr}

\abstract{In this contribution we analyse the heavy-to-light $B$ decays through the Dispersive Matrix method, which can be applied to any semileptonic decays of hadrons once lattice QCD computations of the hadronic Form Factors and of the relevant susceptibilities are available. We will explicitly discuss the application of the Dispersive Matrix approach to both $B \to \pi \ell \nu_\ell$ and $B_s \to K \ell \nu_\ell$ decays. As usual in our analysis strategy, only LQCD computations of the FFs at high values of the momentum transfer will be used to determine the shape of the FFs in the whole kinematical range without making any assumption on their momentum dependence. Then, the experimental data will be used only to obtain our final exclusive determinations of $\vert V_{ub} \vert$. In this way, our calculation of the FFs allows to obtain pure theoretical estimates of several quantities of phenomenological interest, for instance the $\tau/\mu$ ratio of the differential decay rates $R^{\tau/\mu}_{\pi(K)}$, which is an important tool for testing Lepton Flavour Universality. We will also present a summary of all the results obtained so far for semileptonic $B$ decays within the Dispersive Matrix approach.
}

\FullConference{%
  41st International Conference on High Energy physics - ICHEP2022\\
  6-13 July, 2022\\
  Bologna, Italy
}

\begin{document}

\maketitle
\section{Introduction}

Since many years, the heavy-to-light semileptonic $B$-meson transitions are intriguing processes mainly because of a long-standing tension between the inclusive and the exclusive determinations of the CKM matrix element $\vert V_{ub} \vert$. The most recent version of the FLAG report~\cite{FLAG21} quotes for the exclusive estimate of $\vert V_{ub} \vert$ the value $\vert V_{ub} \vert_{excl} \cdot 10^3 = 3.74\,(17)$ from $B \to \pi \ell \nu_\ell$ decays. This value is well compatible with other recent analyses present in literature, see for instance \cite{Biswas:2021qyq}. For what concerns the inclusive determinations, many averages have been recently computed. HFLAV Collaboration~\cite{HFLAV:2022pwe} quotes the value $\vert V_{ub} \vert_{incl} \cdot 10^3 = 4.19 (12)^{(+11)}_{(-12)}$. In the last FLAG review\,\cite{FLAG21} one finds $\vert V_{ub} \vert_{incl} \cdot 10^3 = 4.32\,(29)$. Finally, the last PDG review\,\cite{PDG} quotes $\vert V_{ub} \vert_{incl} \cdot 10^3 = 4.13\,(26)$.

By comparing the FLAG exclusive determination with each of the inclusive values, we find a difference that varies in the $1.5-2\sigma$ range. The present state-of-the-art thus lightens the $\vert V_{ub} \vert$ puzzle with respect to the past years. However, it remains important to address the problem of an accurate determination of $\vert V_{ub} \vert$ from the relevant exclusive channels.

\section{The Dispersion Matrix (DM) approach}
\label{Section2}

For both semileptonic $B \to \pi$ and $B_s \to K$ decays, the differential decay width reads
\begin{equation}
\begin{aligned}
\label{finaldiff333}
&\frac{d\Gamma(B_{(s)}\to \pi(K) \ell \nu)}{dq^2}=\frac{G_F^2 \vert V_{ub} \vert^2}{24\pi^3} \left(1-\frac{m_{\ell}^2}{q^2}\right)^2\\
&\hskip 0.8truecm\times \left[\vert \vec{p}_{\pi(K)}\vert^3 \left(1+\frac{m_{\ell}^2}{2q^2}\right) \vert f_{+}^{\pi(K)} (q^2) \vert^2 + m_{B_{(s)}}^2 \vert \vec{p}_{\pi(K)} \vert \left( 1-\frac{m_{\pi(K)}^2}{m_{B_{(s)}}^2}\right)^2 \frac{3m_{\ell}^2}{8q^2} \vert f_{0}^{\pi(K)} (q^2) \vert^2\right],
\end{aligned}
\end{equation}
where $G_F$ is the Fermi constant, $\vec{p}_{\pi(K)}$ the 3-momentum of the $\pi\, (K)$ meson and $m_{\ell}$ the mass of the produced lepton. From the theoretical point of view, a central role is then played by the hadronic Form Factors (FFs) $f_{+}^{\pi(K)} (q^2),\,f_{0}^{\pi(K)} (q^2)$, which can be directly computed on the lattice with high precision \emph{only} at high values of the momentum transfer $q^2$.

Now, the goal of this proceedings is to describe the FFs entering in Eq.\,(\ref{finaldiff333}) by using the novel Dispersion Matrix (DM) method \cite{DiCarlo:2021dzg}. The DM method, in fact, allows us to study the FFs in the whole kinematical region in a non-perturbative and model-independent way. To be more specific, starting from the available LQCD computations of the FFs at high momentum transfer, we can extrapolate their behaviour in the opposite kinematical region without assuming any functional dependence of the FFs on $q^2$ and using only non-perturbative inputs. 


From the mathematical point of view, the idea is to obtain bounds on the hadronic FFs by using the computations of the FFs on the lattice and of the derivatives of suitable Green functions of bilinear quark operators, that hereafter we will refer to as \emph{susceptibilities}. To be more specific, the susceptibilities follow from the dispersion relation associated to a particular spin-parity quantum channel and have been computed for the first time on the lattice in \cite{paperoV, paperoII} for $b \to u$ and $b \to c$ quark transitions, respectively. Using the DM method, one can obtain from first principles the lower and the upper bounds of a generic FF $F$ for each generic value of $q^2$, $i.e.$
\begin{equation}
F_{\rm lo}(q^2) \leq F(q^2) \leq F_{\rm up}(q^2).
\end{equation}
The explicit expressions of $F_{\rm lo,up}(q^2)$ can be found in \cite{DiCarlo:2021dzg}.


\section{The DM applied to semileptonic $B \to \pi \ell \nu$ decays}

\begin{figure}
\centering
\includegraphics[width=.75\textwidth]{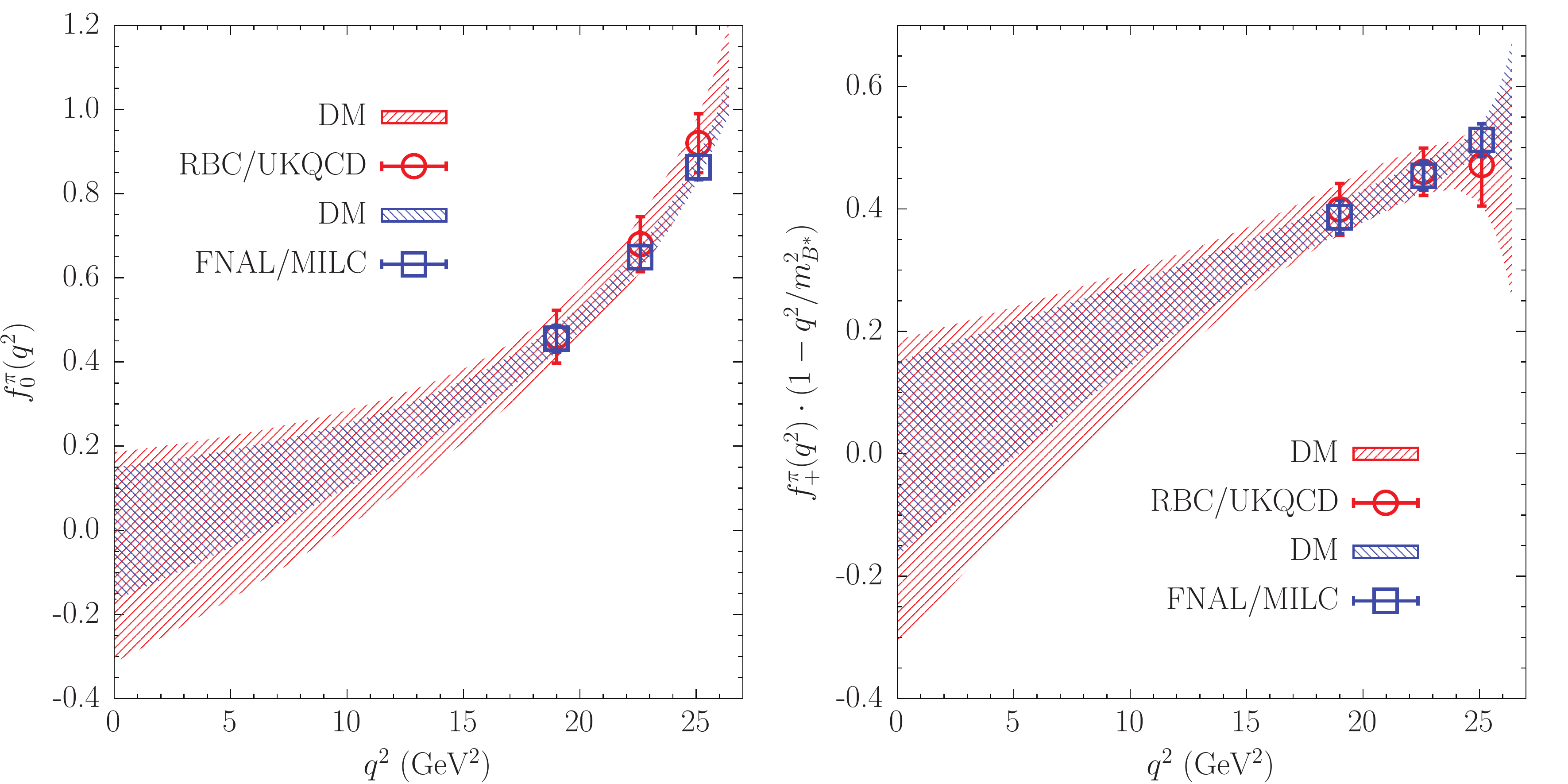}
\caption{\small The scalar $f_0^\pi(q^2)$ (left panel) and vector $f_+^\pi(q^2)$ (right panel) FFs entering the semileptonic $B \to \pi \ell \nu_\ell$ decays computed by the DM method as a function of the 4-momentum transfer $q^2$ using the computations of the FFs by RBC/UKQCD (red points) and FNAL/MILC (blue squares) Collaborations. In the right panel, the vector FF is multiplied by the factor $(1 - q^2 / m_{B^*}^2)$ with $m_{B^*} = 5.325$ GeV.\hspace*{\fill}}
\label{FFsBpi}
\end{figure}

Let us discuss the application of the Dispersive Matrix method to the $B \to \pi \ell \nu$ decays \cite{paperoV}. The FFs $f_+^{\pi}(q^2),\,f_0^{\pi}(q^2)$ have been studied by the RBC/UKQCD\,\cite{Flynn:2015mha} and the FNAL/MILC\,\cite{Lattice:2015tia} Collaborations (the results of a new computation of these hadronic FFs have been recently published also by the JLQCD Collaboration \cite{Colquhoun:2022atw}). The lattice computations of the FFs $f_+^{\pi}(q^2),\,f_0^{\pi}(q^2)$ have been considered in the large-$q^2$ region, namely at $q^2 = \{ 19.0, 22.6, 25.1 \}$ GeV$^2$. In Fig.\,\ref{FFsBpi} we show the red (blue) DM bands coming from the RBC/UKQCD (FNAL/MILC) data, respectively. %


For the extraction of the CKM matrix element we compute bin-per-bin values of $\vert V_{ub} \vert$ for each $q^2$-bin of each available experiment. At present, several experiments \cite{delAmoSanchez:2010af, Ha:2010rf, Lees:2012vv, Sibidanov:2013rkk} have measured the differential branching fractions of the semileptonic $B \to \pi$ transition. All the details of this study can be found in \cite{paperoV}. Our final result for $\vert V_{ub} \vert$ reads
\begin{equation}
\label{VubLASTCOMB}
\vert V_{ub} \vert^{B\pi} \cdot 10^{3}  =  3.62 \pm 0.47.
\end{equation}
Let us mention here that we are currently investigating new strategies to improve our precision on the determination of $\vert V_{ub} \vert$ within the DM approach. Some results can be found in \cite{CKM21}, where our improved determination of the CKM matrix element $\vert V_{ub} \vert$ from semileptonic $B \to \pi$ decays reads 
\begin{equation}
\label{Vubimpr}
\vert V_{ub} \vert^{B\pi}_{\rm impr} \cdot 10^{3}  =  3.88 \pm 0.32.
\end{equation}

Moreover, the DM bands of the FFs $f_{+,0}^{\pi}(q^2)$ can be used to compute fully-theoretical estimates of the $\tau/\mu$ ratio $R^{\tau/\mu}_{\pi} \equiv \Gamma(B \to \pi \tau \nu_\tau)/\Gamma(B \to \pi \mu\nu_\mu)$. Our DM estimate reads $R^{\tau/\mu}_{\pi}=0.793(118)$, which is well compatible with the experimental value $R_{\pi}^{\tau/\mu}\vert_{exp} = 1.05 \pm 0.51$ by Belle \cite{Hamer:2015jsa}.

\section{The DM applied to semileptonic $B_s \to K \ell \nu$ decays}

In the case of the semileptonic $B_s \to K$ transition, the FFs $f_+^{K}(q^2),\,f_0^{K}(q^2)$ have been computed on the lattice by the RBC/UKQCD\,\cite{Flynn:2015mha}, HPQCD\,\cite{Bouchard:2014ypa} and FNAL/MILC\,\cite{Bazavov:2019aom} Collaborations. Also in this case the LQCD values of the FFs have been considered at three values of the momentum transfer in the large-$q^2$ regime, namely $q^2 = \{ 17.6, 20.8, 23.4 \}$ GeV$^2$. We have then implemented the DM method to obtain unitary bands for the FFs (see all the details in \cite{paperoV}).

Our goal is now to determine $\vert V_{ub} \vert$ from semileptonic $B_s \to K$ decays. The LHCb Collaboration has recently observed for the first time these decays\,\cite{Aaij:2020nvo} by measuring the ratio $R_{BF} \equiv \mathcal{B}(B_s^0 \to K^- \mu^+ \nu_{\mu})/\mathcal{B}(B_s^0 \to D_s^- \mu^+ \nu_{\mu})$ in two different $q^2$-bins, namely $q^2 \leq 7\,\mbox{GeV}^2$ and $q^2 \geq 7\,\mbox{GeV}^2$. In order to obtain an exclusive estimate of $\vert V_{ub} \vert$, we made use of the lifetime of the $B_s$-meson, $\tau_{B_s^0} = (1.516 \pm 0.006) \cdot 10^{-12}$ s\,\cite{PDG}, and of the experimental value of the branching ratio $\mathcal{B}(B_s^0 \to D_s^- \mu^+ \nu_{\mu})$ measured by the LHCb Collaboration in\,\cite{Aaij:2020hsi}. Our final result, resulting from a weighted average of the two $q^2$-bins, reads 
\begin{eqnarray}
\label{VubBsKCOMB}
\vert V_{ub} \vert^{B_sK}_{\rm{combined}} \cdot 10^{3} & = & 3.77 \pm 0.48. 
\end{eqnarray}



\begin{figure}[h!]
\centering
\includegraphics[width=.55\textwidth]{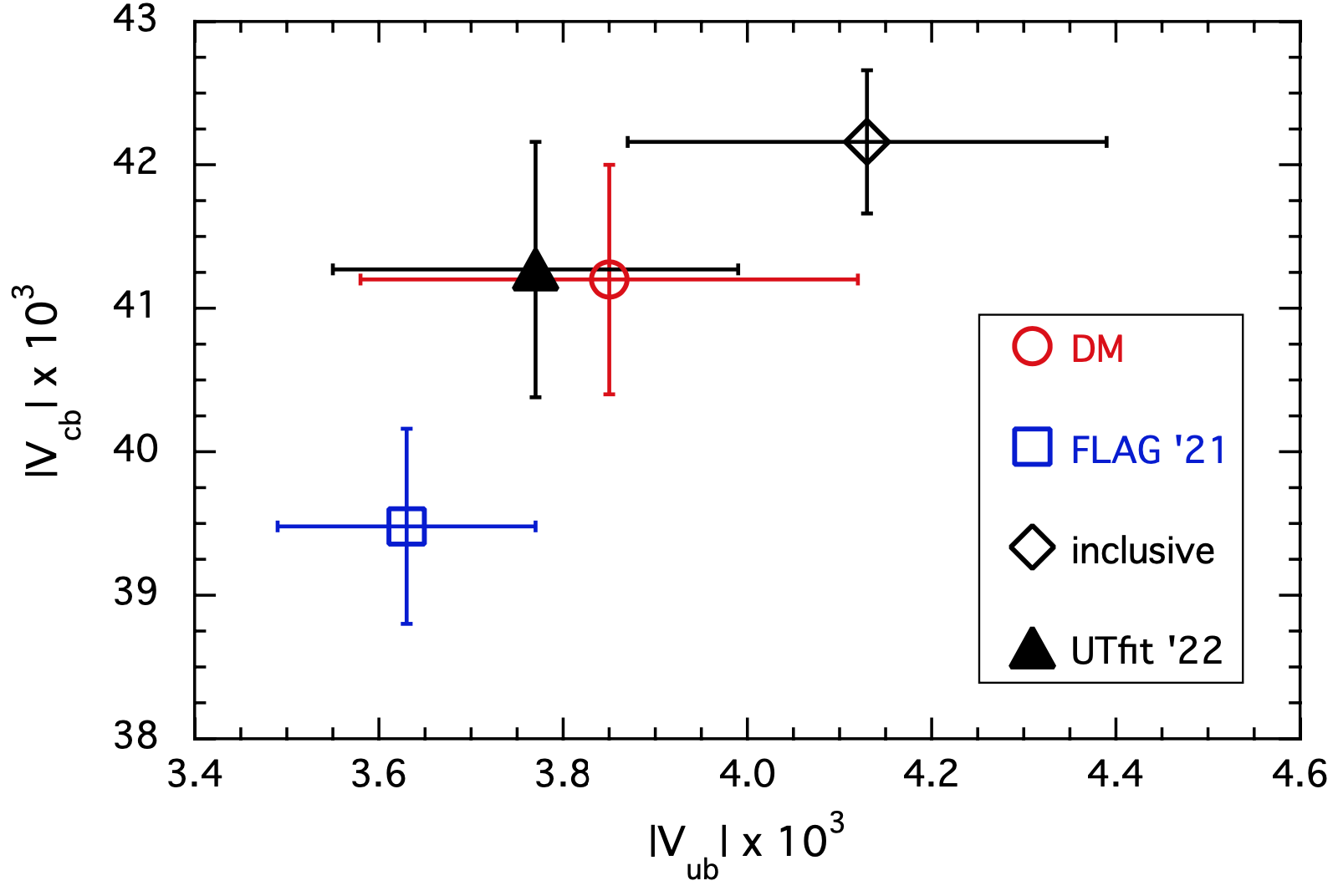}
\caption{\it \small $\vert V_{cb} \vert$ vs $\vert V_{ub} \vert$ correlation plot. The numerical values of the various entries (DM estimates, FLAG 2021 values, inclusive estimates \cite{Bordone:2021oof, PDG}, UTfit determinations \cite{Bona:2022xnf, Bona:2022zhn}) can be found in Table \ref{tab1}.}
\label{Summary}
\end{figure}

\section{Conclusions}

In this contribution we have reviewed the results of the application of the Dispersion Matrix approach to semileptonic $B \to \pi$ and $B_s \to K$ decays \cite{paperoV}. In Figure \ref{Summary} we have condensed the results obtained so far from the application of the DM method, enlarging the discussion also to the semileptonic $B \to D^{(*)}$ \cite{paperoIII, EPJC} and $B_s \to D_s^{(*)}$ \cite{BsDs} transitions. The DM values of the CKM matrix elements represent the averages of all the DM determinations of $\vert V_{cb} \vert$ and $\vert V_{ub} \vert$ from the various decay channels, which are also presented for clarity in Table \ref{tab1}. The value of $\vert V_{ub} \vert$ corresponds to the average of the values in Eqs.\,(\ref{Vubimpr}) and (\ref{VubBsKCOMB}). For both the CKM matrix elements, the DM determinations are compatible with the corresponding inclusive values within the $1\sigma$ level. Furthermore, the DM values are practically identical to the indirect determinations coming for the latest analysis by the UTfit Collaboration \cite{Bona:2022xnf, Bona:2022zhn}. 

\begin{table}[h!]
\centering
\caption{Numerical values of the CKM matrix elements $\vert V_{cb} \vert$ and $\vert V_{ub} \vert$ plotted in Figure \ref{Summary}.}
\label{tab1}       
\begin{tabular}{c|c|c|c|l|cc}
& Decay channel & DM values & FLAG '21 & Inclusive & UTfit '22\\\hline
$\vert V_{cb} \vert \times 10^3$ & $B_{(s)} \to D_{(s)}^{(*)}$ & 41.2 (8) & 39.48 (68) & 42.16 (50) & 41.27 (89) \\\hline
$\vert V_{ub} \vert \times 10^3$ & $B_{(s)} \to \pi(K)$  & 3.85 (27) & 3.63 (14) & 4.13 (26) & 3.77 (22) \\
\end{tabular}
\end{table}

\bibliographystyle{JHEP}
\bibliography{notes_biblio}

\end{document}